\documentclass[11pt,a4paper]{article}
\pdfoutput=1
\usepackage[utf8]{inputenc}
\usepackage{authblk}

\usepackage{tikz}
\usetikzlibrary{shapes.geometric, arrows}
\usepackage{siunitx}
\usepackage{graphicx}
\usepackage{amsmath}
\usepackage{amssymb}
\usepackage{subfig}
\usepackage{wrapfig}
\usepackage{epstopdf}

\usepackage{xpatch}
\xpatchcmd{\author}{\relax#1\relax}{\relax\detokenize{#1}\relax}{}{}

\tikzstyle{left} = [rectangle, minimum width=4.5cm, minimum height=1cm, text centered, text width=4.7cm, draw=black, fill=white!30]
\tikzstyle{right} = [rectangle, minimum width=4cm, minimum height=1cm, text centered, text width=3.8cm, draw=black, fill=white!30]
\tikzstyle{invisible} = [rectangle, minimum width=1cm, minimum height=1cm, text centered, text width=2cm, draw=white, fill=white!30]
\tikzstyle{middle} = [rectangle, minimum width=9cm, minimum height=0.8cm, text centered, text width=9.73cm, draw=black, fill=white!30]
\tikzstyle{arrow} = [thick,->,>=stealth]

\begin{document}

\title{Application of a single-board computer as a low cost pulse generator}
\author[\empty]{Marcus~Fedrizzi\textsuperscript{1,}\thanks{\texttt{marcus.k.fedrizzi@monash.edu}}}
\author[1,2]{Julio~Soria}
\affil[1]{Laboratory for Turbulence Research in Aerospace and Combustion, Department of Mechanical and Aerospace Engineering, \authorcr Monash University, Australia}
\affil[2]{Department of Aeronautical Engineering, King Abdulaziz University, \authorcr Jeddah, Kingdom of Saudi Arabia}
              
\maketitle

\begin{abstract}
A BeagleBone Black (BBB) single-board open-source computer was implemented as a low-cost fully programmable pulse generator. The pulse generator makes use of the BBB Programmable Real-Time Unit (PRU) subsystem to achieve a deterministic temporal resolution of 5 ns, an RMS jitter of 290 ps and a timebase stability on the order of 10 ppm. A python based software framework has also been developed to simplify the usage of the pulse generator.


\end{abstract}
\sloppy
\section{Introduction}
\label{sec:introduction}
Laboratory pulse generators are often used for experiment timing control. Such pulse generators create a series of square waves which use an alternating high and low signal to trigger external devices. One example of their usage is for accurately separating laser pulses in particle image velocimerty (PIV) experiments so that tracer particles can be imaged and the fluid velocity field extracted \cite{raffel07}. The laser pulse emission occurs after receiving the rising or falling edge of the pulse generator signal.

In order to evaluate whether a particular pulse generator is applicable for an experimental measurement, an understanding is needed of how the timing system contributes to the measurement uncertainty and how significant this contribution is. The key information required to assess the contribution of the pulse generator to the measurement uncertainty is its accuracy and precision. 

The accuracy of a pulse generator is a measure of how truly it can produce pulses of a particular period which is limited by the stability of its timebase, often a crystal oscillator. The timebase stability is usually measured in parts-per-million (ppm) or parts-per-billion (ppb) which means for example, a 1 ppm timebase can introduce a timing error up to $\pm$1 ns over 1 ms. The timebase stability of some common commercial pulse generators are 5, 25 and 50 ppm \cite{stanford,bnc,ni} according to their manufacturer specifications. 

The precision of the pulse generator is the consistency of the pulse spacing, which corresponds to the period jitter. Jitter is the deviation of a signal from its ideal signal and the period jitter is the difference of the period produced by the pulse generator, $T_p$, and the ideal period for the set frequency, $T_s = 1/f$  with a numerical measure of the jitter being the root-mean-square (RMS) of at least 1000 samples \cite{jdec_65b}. Two of the commercial systems mentioned previously have also documented RMS jitter of 25 and 50 ps \cite{stanford,bnc}. 

Existing commercial pulse generators can be very accurate and precise however they can be expensive and inflexible. If the pulse generator is not a significant source of experimental uncertainty, a lower precision and accuracy can be justified and a lower cost alternative considered. Commercial pulse generators can also be limited in the number of inputs and outputs and their adaptability to complex experiments. It is sometimes preferable to use a fully programmable system for experiments which have unusual timing requirements or for experiments that are difficult or impossible to implement using a typical pulse generator. 

Some fully programmable timing control systems have previously been implemented using computers running real-time operating systems \cite{soria03}. This system utilised the parallel port of the host computer to send pulses through an isolating device to where they are needed. With the parallel port now an obsolete feature on off the shelf computers and the inherent latency and jitter issues with real-time operating systems, which limits their response times and maximum operating frequencies \cite{aroca09}, it is appropriate to consider other alternatives.

\begin{figure}
\centering
\includegraphics[width=0.65\textwidth]{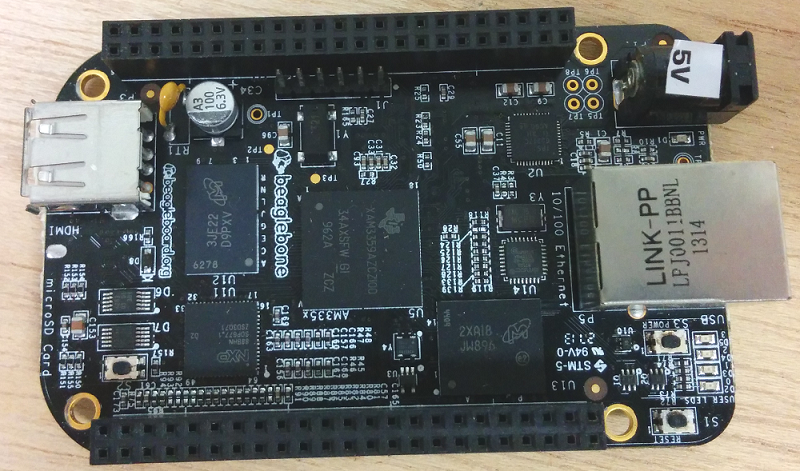}
\caption{The BeagleBone Black, a credit card sized, low-cost single-board computer.}
\label{fig:bbb}
\end{figure}

Recent advances in mobile technology and microcontrollers have led to the development of many low cost systems including some more popular devices such as the Arduino, Raspberry Pi and BeagleBone Black (BBB). The BBB, shown in Figure \ref{fig:bbb} was chosen for this work because it has the unique feature amongst these development platforms of being a complete linux computer with two dedicated real-time processors.

\section{Implementation}
\label{sec:implementation}
The BBB is a low-cost, credit card sized single-board computer based on the TI Sitara AM335x ARM Cortex-A8 system-on-chip. This system contains a main 1 GHz processor as well as a Programmable Realtime Unit (PRU) subsystem. This subsystem consists of two processors, the PRUs, which operate independently from the main processor. One or both of the PRUs can be dedicated to sending or receiving timing signals which avoids the software related jitter associated with real-time operating systems. The two PRU processors have a clock frequency of \SI{200}{\mega\hertz} and interface directly with up to twenty-six input and output pins which can be used for control purposes. Each output pin can be toggled in one clock cycle, giving the BBB a maximum temporal resolution of \SI{5}{\nano\second}. 

The BBB was used in its out of the box state with the addition of PyPruss \cite{pypruss} which is freely available open-source software to operate the PRU subsystem using python. The PRU code was written in assembly language \cite{ti_assembly} where each instruction requires one processor clock cycle to execute. Switching the output pins requires changing a particular bit in register 30 on the PRU while inputs can be read from register 31.

The major task for the board to achieve was to repetitively produce pulses at a set frequency with defined temporal offsets from each other and with a fixed pulse width. This involves switching a particular pin to high, waiting a given amount of time and switching it back to low. This  procedure is simple for a single pin however it leads to some tedious ordering of switching tasks and calculating delay times when dealing with multiple signals which have different offsets, especially when pulses overlap.

\begin{figure}
\centering
\begin{tikzpicture}[node distance=2cm]
\node (start) [left] {Set up control loops. Specify which outputs are to be used and the end condition for the loop.};    
\node (user0) [left, below of=start] {Define signal offsets and pulse widths};
\node (setup) [right, right of=user0, xshift=3.5cm,yshift=0.3cm] {Select which PRU to run on and map pins to outputs};
\node (pro1) [left, below of=user0] {Order switching tasks and calculate delay times};
\node (pro2) [right, right of=pro1, xshift=3.5cm] {Write and compile assembly language code for the PRU};
\node (setup2) [right,below of=pro2] {Configure pins and enable PRU subsystem};
\node (pro4) [left, below of=pro1] {Send compiled code to PRU};
\node (wait) [middle, below of=pro4, xshift=2.52cm] {Wait for code to finish or user interrupt and disable the PRU};
\node (note) [invisible, right of=start, xshift=3.5cm,yshift=0.1cm] {USER INPUT};
\draw[thick,dashed] (-2.7,1.2) -- (-2.7,-2.71) -- (7.7,-2.71)  -- (7.7,1.2) -- (-2.7,1.2);
\draw [arrow] (start) -- (user0);
\draw [arrow] (user0) -- (pro1);
\draw [arrow] (pro1) -- (pro2);
\draw [arrow] (setup) -- (pro2);
\draw [arrow] (pro4) -- (0,-7.46);
\draw [arrow] (pro2) -- (setup2);
\draw [arrow] (setup2) -- (pro4);
\end{tikzpicture}
\caption{Implementation of the python framework for developing experiment control programs.}

\label{fig:program_flow}
\end{figure}
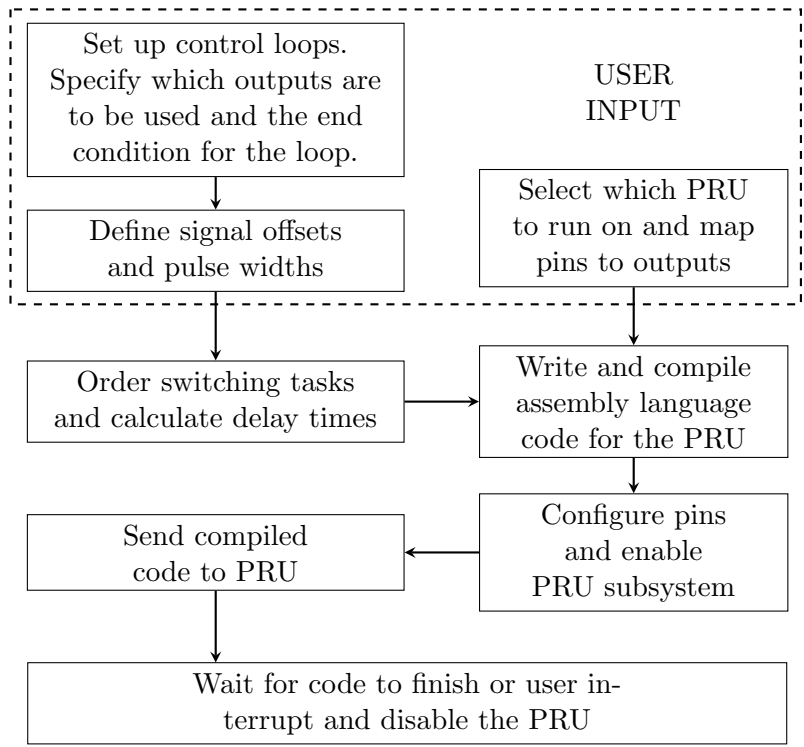

A python-based framework was developed to order the switching tasks and calculate the delays between them as shown in the flow chart in Figure \ref{fig:program_flow}. This framework shields the user from the specifics of writing, compiling and executing the PRU assembly code. Behind this layer of abstraction, the user sets up one or more control loops in a python script, specifying which output pins will be used in the loop and whether to repeat the loop a finite number of times or indefinitely until interrupted by the user. Multiple loops can be linked for burst mode functionality, frequency scaling or more complex triggering. Within the control loops the outputs are given a pulse width and an offset time. 

Once the user defined parameters are set, the switching tasks are interpreted in each loop and sorted based on their execution time. An appropriate delay time between tasks is calculated and the assembly language code is produced and compiled. The pins on the board are configured to their required mode and the code is sent to the PRUs for execution. 

\begin{figure}
\centering
\includegraphics[width=0.65\textwidth]{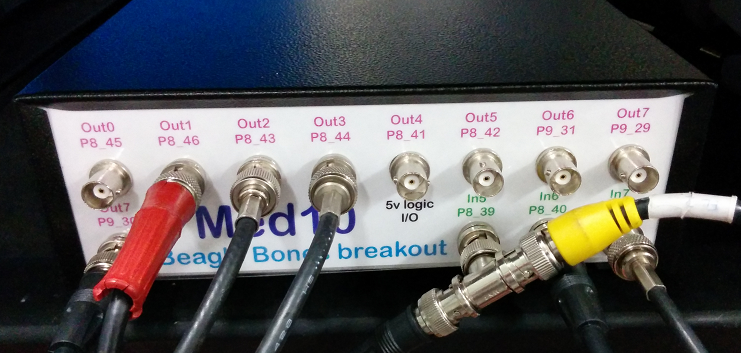}
\caption{The BeagleBone Black pulse generator front panel. This implementation has nine outputs and three inputs.}
\label{fig:bbb_box}
\end{figure}

The output timing pulses were translated from the 3.3 V BBB system voltage to a 5 V TTL level signals and was isolated using radio-frequency (RF) isolators. These isolators are able to rapidly respond to signals within 10 ns with low jitter and were powered by an external 5V power supply connected to a common ground. The BBB and isolator board were enclosed in an aluminium case to shield it from external electrical interference with signals being sent and received through BNC cables, shown in Figure \ref{fig:bbb_box}. The driving software is freely available for academic use on the LTRAC website \cite{ltrac_web}. A component list and wiring schematic is also provided for a nine output, three input channel box costing approximately 100 AUD, which is between 2-5\% of the online prices of the commercial pulse generators. 

\section{Performance}
\begin{figure}
\centering
\includegraphics[width=0.9\textwidth]{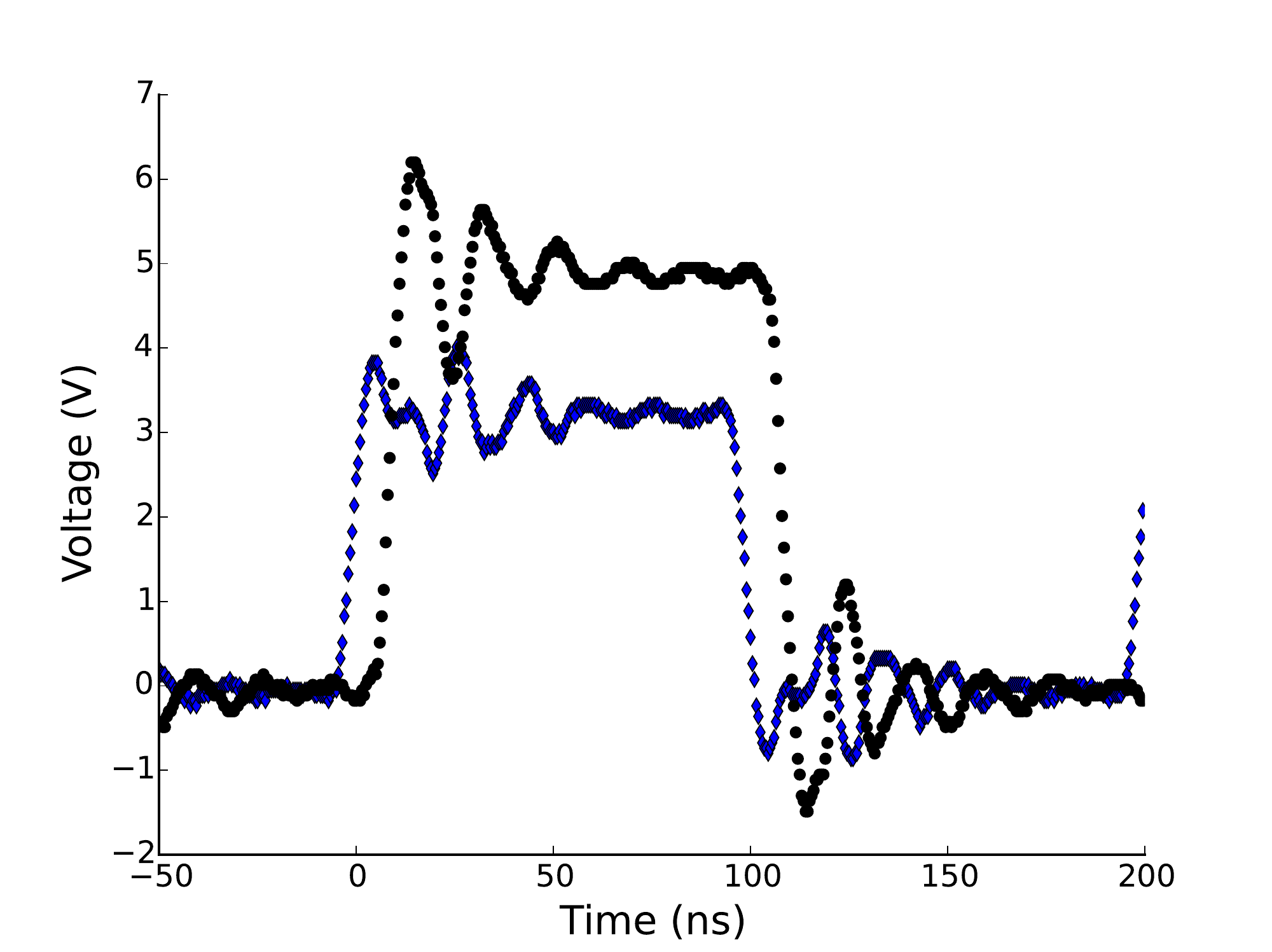}
\caption{The output of a 5 MHz, 100 ns pulse from the BBB (blue diamonds) and the isolated output of the pulse generator (black circles)}
\label{fig:pulse_shape}
\end{figure}

The performance of the BBB pulse generator was assessed using a 2 GHz digital storage oscilloscope (DSO) with a 15 ppm timebase stability. An example of an output pulse measured by the DSO directly from the BBB pin and after isolation is shown in Figure \ref{fig:pulse_shape} for a 5 MHz signal at 50\% duty cycle. The mean rise time of the isolated pulse was 4.6 ns, defined as the time taken for the signal to rise from 0.5 V to 4.5 V. The rise time could influence the triggering of different devices which do not switch at the exact same voltage.

\begin{figure}
\centering
\includegraphics[width=0.9\textwidth]{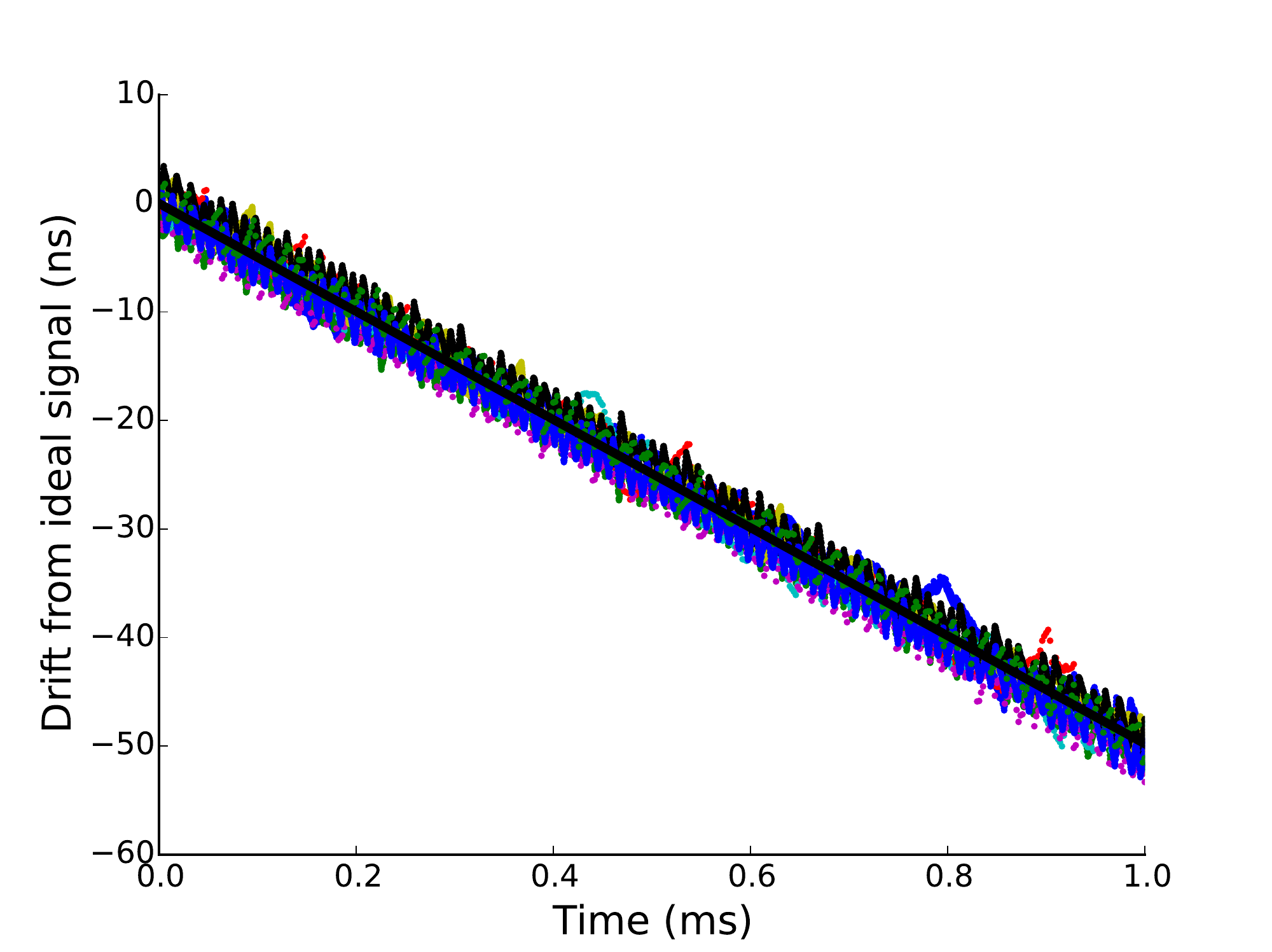}
\caption{The deviation of the signal from the ideal pulse time for nine different datasets. A linear least squares fit is shown as the black line and the gradient is interpreted as the timebase stability.}
\label{fig:jitter_drift}
\end{figure}

The BBB uses a reference oscillator as an input into a phase locked loop (PLL) which generates the 200 MHz PRU clock frequency. The jitter in the PRU clock is a function of jitter in the reference oscillator however information on the BBB oscillator stability is not readily available. The timing accuracy was instead assessed by taking several 1 ms pulse sequences from the DSO. The resultant deviation from the ideal time was calculated by taking the time of the $n$th rising edge, $t_n$ of the signal by linearly interpolating the DSO output to find the time it passed a threshold voltage, then subtract the ideal edge time, $nT_s$. The deviation is shown in Figure \ref{fig:jitter_drift} and a least squares fit is indicated by the black solid line. The slope of this graph is interpreted as the timebase stability, corresponding to approximately 50 ppm. Given the uncertainty in the timebase of the DSO, this measurement has an uncertainty of $\pm$ 15 ppm, but even with this uncertainty, it is still of the same order of magnitude as two of the three commercial systems considered.

\begin{figure}
\centering
\includegraphics[width=0.9\textwidth]{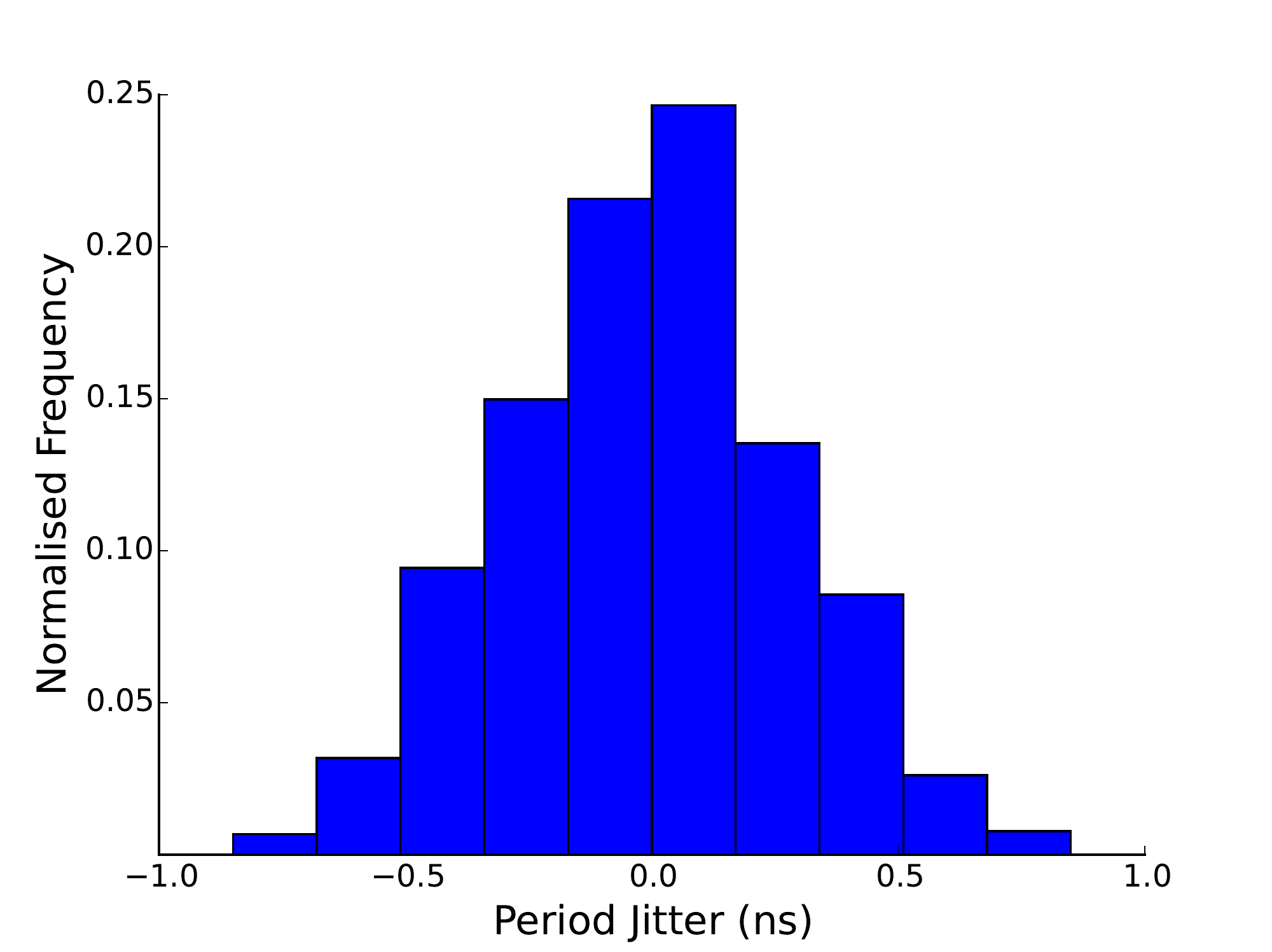}
\caption{Histogram of the period jitter of a 5 MHz output signal.}
\label{fig:jitter_hist}
\end{figure}

The jitter in the system was found by exporting 10000 5 MHz, 100 ns pulses from the DSO and finding $T_p$ by calculating the time difference between successive rising edges using linear interpolation. The histogram of the period jitter $T_p - T_s$ is shown in Figure \ref{fig:jitter_hist}. The maximum and minimum values of the period jitter were within $\pm$850 ps of $T_s$, and the RMS of the period jitter was 290 ps.

\begin{figure}
\centering
\includegraphics[width=0.9\textwidth]{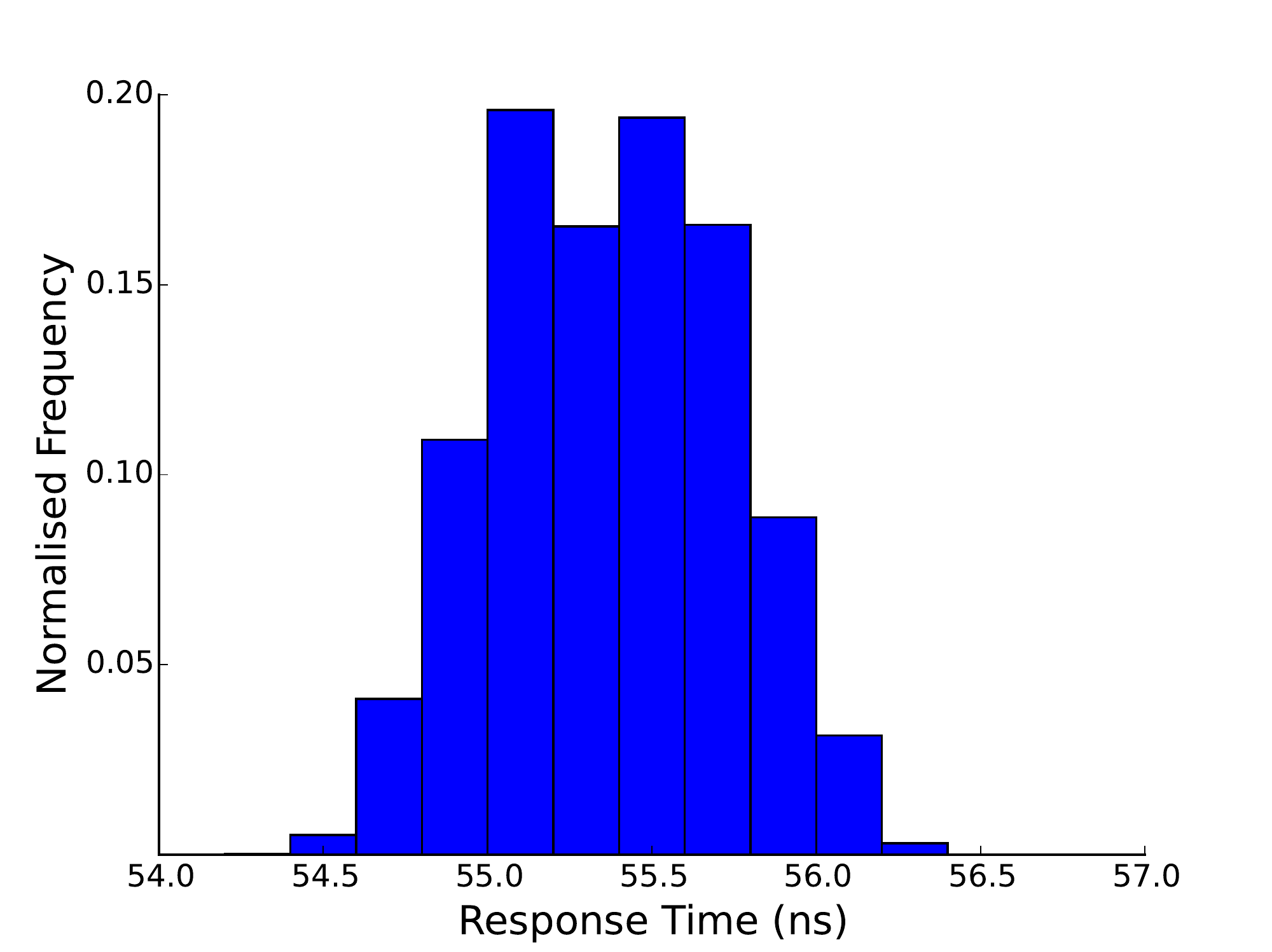}
\caption{Histogram of the time for the BBB pulse generator to respond to an input}
\label{fig:delay_histogram}
\end{figure}

The RF isolators introduce a delay as seen in Figure \ref{fig:pulse_shape} which increases the time for the BBB to respond to an input, in addition to the time normally taken for a BBB to detect an input and provide an output response. The time delay between the leading edge of 10000 input pulses and the leading edge of the corresponding outputs were measured and the result is shown in Figure \ref{fig:delay_histogram}. The mean delay in responding to an input signal was 55.4 ns with an RMS jitter of 350 ps and minimum and maximum response time of 54.4 and 56.3 ns respectively. Using an input, the BBB could be triggered using an external clock signal with a more stable timebase to improve the accuracy of the pulse generator. 

The pulse generator is most useful in not only generating pulses at a specific frequency but also because it can specify delays between many different channels. One limitation is that pin switching from the PRU is sequential and two pins cannot be switched simultaneously so there is a 5 ns offset between any two pins switched by a single PRU. The specifications of the PRU indicate that there is a maximum internal skew of 1 ns and 5 ns between pins switched by PRU0 and PRU1 respectively \cite{tiam335x_datasheet}. This skew value limits the accuracy that can be achieved in the delay between different pins.

\section{Conclusion}
The BBB pulse generator is a promising experiment timing control system with a timebase stability in the order of 10 ppm, which is the same order of magnitude as some commercial systems. The RMS jitter is 290 ps which is an order of magnitude larger than the manufacturer specifications for some commercial systems however it is several orders of magnitude smaller than a real-time operating system. While commercial pulse generators out-perform the BBB pulse generator in accuracy and precision, the BBB has a large number of inputs and outputs and is fully programmable and the component cost for this implementation is 2-5\% of the considered commercial systems. This allows it to be used instead of real-time operating system pulse generator implementations and for experiments where commercial pulse generators cannot provide enough inputs and outputs or sufficient flexibility to control the experimental equipment.

\label{sec:conclusion}

\section*{Acknowledgements}
This work was funded by the Defence Science Technology Organisation, the Australian Research Council and an Australian Postgraduate Award.    

\bibliographystyle{spphys}      
\raggedright
\bibliography{ref}

\begin{thebibliography}{10}
\providecommand{\url}[1]{{#1}}
\providecommand{\urlprefix}{URL: }
\expandafter\ifx\csname urlstyle\endcsname\relax
  \providecommand{\doi}[1]{DOI \discretionary{}{}{}#1}\else
  \providecommand{\doi}{DOI \discretionary{}{}{}\begingroup
  \urlstyle{rm}\Url}\fi

\bibitem{raffel07}
M.~Raffel, C.~Willert, S.~Wereley, J.~Kompenhans, \emph{Particle Image
  Velocimetry: A Practical Guide} (Springer, Berlin, 2007)

\bibitem{stanford}
{Stanford Research Systems}.
\newblock {DG645} digital delay/pulse generator.
\newblock
  \urlprefix\url{http://www.thinksrs.com/downloads/PDFs/Catalog/DG645c.pdf}.
\newblock Accessed: 09-04-2015

\bibitem{bnc}
{Berkeley Nucleonics Corporation}.
\newblock Model 575 digital delay / pulse generator.
\newblock
  \urlprefix\url{\url{http://www.berkeleynucleonics.com/products/model\_575.html}}.
\newblock Accessed: 09-04-2015

\bibitem{ni}
{National Instruments}.
\newblock {NI} {PCI}-6602.
\newblock \urlprefix\url{http://sine.ni.com/nips/cds/view/p/lang/en/nid/1123}.
\newblock Accessed: 09-04-2015

\bibitem{jdec_65b}
{JEDEC Solid State Technology Association}.
\newblock {JEDEC} standard {JESD65B} (2003)

\bibitem{soria03}
J.~Soria, T.~New, T.~Lim, K.~Parker, Exp Thermal Fluid Sci \textbf{27}(5), 667
  (2003)

\bibitem{aroca09}
R.V. Aroca, G.~Caurin, Workshop de Sistemas Operacionais  (2009)

\bibitem{pypruss}
E.~Bakken.
\newblock intelligentagent / pypruss - bitbucket (2014).
\newblock \urlprefix\url{https://bitbucket.org/intelligentagent/pypruss}.
\newblock Accessed: 09-04-2015

\bibitem{ti_assembly}
{Texas Instruments}.
\newblock Pru assembly instructions.
\newblock
  \urlprefix\url{http://processors.wiki.ti.com/index.php/PRU\_Assembly\_Instructions}.
\newblock Accessed: 09-04-2015

\bibitem{ltrac_web}
{Laboratory for Turbulence Research in Aerospace and Combustion (LTRAC)}.
\newblock Beaglebone pulse/delay generator.
\newblock \urlprefix\url{https://ltrac.eng.monash.edu.au/software}

\bibitem{tiam335x_datasheet}
{Texas Instruments}.
\newblock {AM335x Sitara}™ {Processors}.
\newblock \urlprefix\url{\url{http://www.ti.com/lit/ds/symlink/am3357.pdf}}.
\newblock Accessed: 09-04-2015

\end{thebibliography}

\end{document}